# Electron Mobility in Bulk *n*-Doped SiC-Polytypes 3*C*-SiC, 4*H*-SiC, and 6*H*-SiC: A Comparison

**C. G. Rodrigues***

*School of Exact Sciences and Computing, Pontifical Catholic University of Goiás, CP 86, Goiânia, Goiás, 74605-010 Brazil*

**e-mail: cloves@pucgoias.edu.br*

**Abstract**—This communication presents a comparative study on the charge transport (in transient and steady state) in bulk *n*-type doped SiC-polytypes: 3*C*-SiC, 4*H*-SiC and 6*H*-SiC. The time evolution of the basic macrovariables: the "electron drift velocity" and the "non-equilibrium temperature" are obtained theoretically by using a Non-Equilibrium Quantum Kinetic Theory, derived from the method of Nonequilibrium Statistical Operator (NSO). The dependence on the intensity and orientation of the applied electric field of this macrovariables and mobility are derived and analyzed. From the results obtained in this paper, the most attractive of these semiconductors for applications requiring greater electronic mobility is the polytype 4*H*-SiC with the electric field applied perpendicular to the *c*-axis.

## 1. INTRODUCTION

The intrinsic properties of the wide band gap semiconductors, specifically the wide band gap energy that enables higher junction operating temperatures, make these materials particularly attractive for high power device applications. A very attractive material is the semiconductor Silicon Carbide (SiC). The SiC is a wide-bandgap semiconductor with high saturated carrier drift velocity, small dielectric constant, high thermal conductivity, being an attractive semiconductor for high power device applications [1–8]. Silicon carbide can form in many distinct crystal structures (know as polytypes), with some of the most common being 3*C*-SiC (cubic), 4*H*-SiC (hexagonal) and 6*H*-SiC (hexagonal). 3*C*-SiC is the only possible cubic polytype. The 4*H*-SiC and 6*H*-SiC have anisotropic transport properties, because the electron effective masses to differ notably between orientation along the basal plane perpendicular to the *c*-axis (which we labeled $m^*_{e\perp}$) and that parallel to the *c*-axis (which we labeled $m^*_{e\|}$). We draw attention to the fact that the degree and characteristics anisotropy are different for each polytype [1].

The optical and transport properties of semiconductors have been studied by using Nonequilibrium Green's Functions Techniques [9], Monte Carlo simulation [10], balance equation approach [11], Boltzmann transport equations [12], etc. In this paper, we use the "*Nonequilibrium Statistical Operator*" (NSO) [13–22]. The NSO is a powerful formalism that seems to offer an elegant and concise way for an analytical treatment in the theory of irreversible processes, adequate to deal with a large class of experimental situations, and physically clear picture of irreversible processes. The NSO invented by D.N. Zubarev is also practical and efficient in the study of the optical [23, 24] and carrier dynamics in semiconductors [25–31]. More specifically, a Non-Equilibrium Quantum Kinetic Theory [32] derived from NSO was used in this paper.

In this work NSO has been used for a comparative study of bulk nonlinear charge transport in *n*-type doped SiC-polytypes: 3*C*-SiC, 4*H*-SiC and 6*H*-SiC; for the case of polytypes 4*H*-SiC and 6*H*-SiC the charge transport is analyzed when the transport direction is along the *c*-axis, or when the transport direction is in the plane perpendicular to it. We obtain theoretically the drift velocity of electrons, $v_e(t)$, and the non-equilibrium temperature of electrons, $T^*_e(t)$, in two regimes: the transient state and steady state. The dependence of these two macrovariables, $v_e(t)$ and $T^*_e(t)$, on the electric field $\mathscr{E}$ applied in the orientation parallel (which we labeled $\mathscr{E}_\|$) or perpendicular (which we labeled $\mathscr{E}_\perp$) to the *c*-axis is derived and analyzed.

We emphasize that to develop high performance electronic devices, beyond optimizing the fabrication steps, a good knowledge of the transport properties is required [33]. As an example, the electron mobility is a very important property, affecting the device perfor-mances and, hence, can be considered as a figure of merit of the microscopic quality of the epilayers.

## 2. THEORETICAL FRAMEWORK

The Hamiltonian of the $n$-type doped semiconductor is taken as composed of: the energy of the free phonons, the energy of the free electrons, the interaction of the electrons with the constant electric field, the interaction of the electrons with the impurities, the electron–phonon interaction, the anharmonic interaction of the LO-phonons with AC-phonons, the interaction of the AC-phonons with the thermal bath (external thermal reservoir at temperature $T_0$). Regarding the applied constant electric field, we emphasize that the electrical field is switched on at time zero which after that has a constant value so that we observe the initial transient state before the stationary state is established.

The nonequilibrium thermodynamic state of the system is very well characterized for the set of *basic variables*:

$$\{E_e(t), N, \mathbf{P}_e(t), E_{\mathrm{LO}}(t), E_{\mathrm{AC}}\}, \quad (1)$$

that is: the energy of the electrons $E_e(t)$; the number $N$ of the electrons; the linear momentum $\mathbf{P}_e(t)$ of the electrons; the energy of the LO-phonons $E_{\mathrm{LO}}(t)$; and the energy of the AC-phonons $E_{\mathrm{AC}}(t)$.

The *nonequilibrium thermodynamic variables* associated with variables of set (1), are [16, 17, 34]

$$\{\beta_e^*(t), -\beta_e^*(t)\mu_e^*(t), -\beta_e^*(t)\mathbf{v}_e(t), \beta_{\mathrm{LO}}^*(t), \beta_{\mathrm{AC}}^*(t)\}, \quad (2)$$

where $\mathbf{v}_e(t)$ is the drift velocity of the electrons, $\mu_e^*(t)$ is the quasi-chemical potential and

$$\beta_e^*(t) = \frac{1}{k_{\mathrm{B}}T_e^*(t)}, \quad (3)$$

$$\beta_{\mathrm{LO}}^*(t) = \frac{1}{k_{\mathrm{B}}T_{\mathrm{LO}}^*(t)}, \quad (4)$$

$$\beta_{\mathrm{AC}}^*(t) = \frac{1}{k_{\mathrm{B}}T_{\mathrm{AC}}^*(t)}, \quad (5)$$

where $k_{\mathrm{B}}$ is Boltzmann constant, $T_e^*(t)$ is the nonequilibrium temperature of the electrons, $T_{\mathrm{LO}}^*(t)$ is the non-equilibrium temperature of the LO-phonons, and $T_{\mathrm{AC}}^*(t)$ is the nonequilibrium temperature of the AC-phonons [34–36].

By using the Non-Equilibrium Quantum Kinetic Theory based on the NSO, we obtain the equations of evolution for the basic macrovariables [32]:

$$\frac{d\mathbf{P}_e(t)}{dt} = -nVe\mathscr{E} + \mathbf{J}_{\mathbf{P}_{\mathrm{ph}}}(t) + \mathbf{J}_{\mathbf{P}_{\mathrm{imp}}}(t), \quad (6)$$

$$\frac{dE_e(t)}{dt} = -\frac{e}{m_e^*}\mathscr{E}\cdot\mathbf{P}_e(t) + J_{E,\mathrm{ph}}(t), \quad (7)$$

$$\frac{dE_{\mathrm{LO}}(t)}{dt} = J_{\mathrm{LO}}(t) - J_{\mathrm{LO},an}(t), \quad (8)$$

$$\frac{dE_{\mathrm{AC}}(t)}{dt} = J_{\mathrm{AC}}(t) + J_{\mathrm{LO},an}(t) - J_{\mathrm{AC},dif}(t), \quad (9)$$

where $n$ is the concentration of electrons fixed by doping ($n = N/V$); $e$ is the elementary charge; $\mathscr{E}$ is the constant applied electric field; $E_e(t)$ is the energy of the electrons; $m_e^*$ is electron effective mass; $\mathbf{P}_e(t)$ is the linear momentum of the electrons; $E_{\mathrm{LO}}(t)$ is the energy of the longitudinal optical phonons that interact with the electrons via optical deformation potential and Fröhlich potential; $E_{\mathrm{AC}}(t)$ is the energy of the acoustic phonons that interact with the electrons via acoustic deformation potential. We emphasize that for the polytypes 4$H$-SiC and 6$H$-SiC $\mathscr{E}$ is the constant electric field applied perpendicular ($\mathscr{E}_\perp$) or parallel ($\mathscr{E}_\parallel$) to the $c$-axis direction. Moreover, in Eqs. (6) to (9), $m_e^* = m_{e\parallel}^*$ when the constant electric field is applied parallel to the $c$-axis, or $m_e^* = m_{e\perp}^*$ when the constant electric field is applied perpendicular to the $c$-axis.

In Eq. (6), the first right hand term ($-nVe\mathscr{E}$) is the driving force created by the constant applied electric field. The second term, $\mathbf{J}_{\mathbf{P}_{\mathrm{ph}}}(t)$, is the rate of electron momentum transfer due to interaction with the LO and AC phonons. The third term, $\mathbf{J}_{\mathbf{P}_{\mathrm{imp}}}(t)$, is the rate of electron momentum transfer due to interaction with the ionized impurities [37]. In Eq. (7), the first right hand term ($-e\mathscr{E}\cdot\mathbf{P}_e(t)/m_e^*$) is the rate of energy transferred from the constant applied electric field to the electrons, and the second term, $J_{E,\mathrm{ph}}(t)$, is the transfer of the energy of the electrons to the LO and AC phonons.

In Eqs. (8) and (9), the first right hand term accounts for the rate of change of the energy of the LO or AC phonons, respectively, due to interaction with the electrons, that is, the gain of energy transferred to the phonons from the hot electrons. Like this, the sum of contributions $J_{\mathrm{LO}}(t)$ and $J_{\mathrm{AC}}(t)$ is equal to the last term in Eq. (7), $J_{E,\mathrm{ph}}(t)$, but with change of sign. The second term in Eq. (8), $J_{\mathrm{LO},an}(t)$, is the rate of transfer of energy from the LO-phonons to the AC-phonons via anharmonic interaction. We noticed that the therm $J_{\mathrm{LO},an}(t)$ is the same, however with different sign in Eqs. (8) and (9). Shutting down our analysis of Eqs. (6) to (9), we noticed that the last term in Eq. (9), $J_{\mathrm{AC},dif}(t)$, is the diffusion of heat from the AC-phonons to the thermal bath at temperature $T_0$. The detailed expressions for the collision operators are given in Appendix.

We emphasize that to close the system of equations of evolution we must establish the relationship between the basic variables, set Eq. (1), and the *non-* *equilibrium thermodynamic variables*, set Eq. (2). These relationships are:

$$E_e(t) = \sum_{\mathbf{k}} \epsilon_k f_k(t) = nV\left[\frac{3}{2}k_B T_e^*(t) + \frac{1}{2}m_e^* v_e(t)^2\right], \quad (10)$$

$$\mathbf{P}_e(t) = \sum_{\mathbf{k}} \hbar\mathbf{k} f_k(t) = nVm_e^*\mathbf{v}_e(t), \quad (11)$$

$$E_{\rm LO}(t) = \sum_{\mathbf{q}} \hbar\omega_{\mathbf{q},\rm LO}\nu_{q,\rm LO}(t), \quad (12)$$

$$E_{\rm AC}(t) = \sum_{\mathbf{q}} \hbar\omega_{\mathbf{q},\rm AC}\nu_{q,\rm AC}(t), \quad (13)$$

which are known as "*nonequilibrium thermodynamic equations of state*" [16, 17, 34].

In a closed calculation we obtain to phonon populations

$$\nu_{\mathbf{q},\rm LO}(t) = \frac{1}{e^{\beta_{\rm LO}^*(t)\hbar\omega_{\mathbf{q},\rm LO}} - 1}, \quad (14)$$

$$\nu_{\mathbf{q},\rm AC}(t) = \frac{1}{e^{\beta_{\rm AC}^*(t)\hbar\omega_{\mathbf{q},\rm AC}} - 1}, \quad (15)$$

and for to electron populations (in the nondegenerate state)

$$f_{\mathbf{k}}(t) = \exp\{-\beta_e^*(t)[\epsilon_{\mathbf{k}} - \hbar\mathbf{k}\cdot\mathbf{v}_e(t) - \mu_e^*(t)]\}, \quad (16)$$

where

$$\mu_e^*(t) = \frac{1}{\beta_e^*(t)}\ln\left\{\frac{4n\hbar^3}{e^{\beta_e^*(t)m_e^*v_e(t)^2/2}}\sqrt{\left(\frac{\pi\beta_e^*(t)}{2m_e^*}\right)^3}\right\}. \quad (17)$$

We emphasize that in Eqs. (12) and (14) we take the "Einstein model" dispersionless frequency relation for LO phonons, that is: $\hbar\omega_{\mathbf{q},\rm LO} = \hbar\omega_0$ (i.e. all LO phonons have the same angular frequency $\omega_0$), and in Eqs. (13) and (15) we take the "Debye model" dispersionless frequency relation for AC phonons, that is: $\hbar\omega_{\mathbf{q},\rm AC} = \hbar q v_s$, where $v_s$ is the sound velocity. We noticed that the AC and LO phonons were considered to be internally thermalized, disregarding possible differentiated distribution of populations in reciprocal space as a result of what can be termed as Fröhlich–Cherenkov effect [38–41]. The inhomogeneous distribution in reciprocal space is restrict to a very small region of the Brillouin zone [38–41] and such effect can be neglected in the present study.

In next section we perform the numerical calculations for the SiC-polytypes. For this, we have used the parameters for polytypes SiC listed in Table 1. We notice that in Table 1 $m_e^*$ is electron effective mass (in units of $m_0$, the free electron mass); $m_e^* = m_{e\|}^*$ is electron effective mass when the constant electric field is applied parallel to the $c$-axis, and $m_e^* = m_{e\perp}^*$ is electron effective mass when the constant electric field is applied perpendicular to the $c$-axis (in units of $m_0$); $a$ and $c$ are lattice constants (in Å); $\hbar\omega_{\rm LO}$ is longitudinal optical phonon energy (in eV); ρ is mass density (in g/cm$^3$); $v_{sl}$ is longitudinal sound velocity and $v_{st}$ is transversal sound velocity (in $\times 10^5$ cm/s); $E_1$ is acoustic deformation potential (in eV); $\epsilon_0$ is low frequency dielectric constant; $\epsilon_\infty$ is high frequency dielectric constant.

**Table 1.** Parameters of polytypes SiC

| | 3*C*-SiC | 4*H*-SiC | 6*H*-SiC |
|---|---|---|---|
| $m_e^*$ | 0.346 [42] | – | – |
| $m_{e\|}^*$ | – | 0.22 [47] | 0.34 [51] |
| $m_{e\perp}^*$ | – | 0.18 [47] | 0.24 [51] |
| $a$ | 4.36 [1] | 3.073 [48] | 3.080 [48] |
| $c$ | – | 10.053 [48] | 15.120 [48] |
| $\hbar\omega_{\rm LO}$ | 0.12 [43] | 0.12 [49] | 0.12 [49] |
| $\rho$ | 3.21 [44] | 3.12 [49] | 3.2 [49] |
| $v_{sl}$ | 11.2 [43] | 13.5 [50] | 13.5 [49] |
| $v_{st}$ | 6.44 [43] | 7.50 [50] | 7.50 [49] |
| $E_1$ | 22 [45] | 11.6 [49] | 11.2 [52] |
| $\epsilon_0$ | 9.72 [46] | 10.0 [50] | 9.7 [49] |
| $\epsilon_\infty$ | 6.52 [46] | 6.7 [50] | 6.5 [49] |

## 3. RESULTS AND DISCUSSION

The time evolution of the basic intensive nonequilibrium thermodynamic variables are obtained after numerically solve the set of coupled nonlinear integro-differential equations: Eq. (6) to Eq. (9). We assumed that the reservoir temperature is 300 K, and an electron concentration $n$ equal to the ionized dopant concentration $n_I$, that is, $n = n_I = 10^{16}$ cm$^{-3}$. For the polytypes 4*H*-SiC and 6*H*-SiC the electric field is applied perpendicular to the $c$-axis ($\mathscr{E}_\perp$) or parallel to the $c$-axis ($\mathscr{E}_\|$).

Figures 1 and 2 show, respectively, the non-equilibrium temperature of electrons, $T_e^*(t)$, and drift velocity of electrons, $v_e(t)$, in function of time (in picoseconds) for an electric field strength of 80 kV/cm (Fig. 1a) and 20 kV/cm (Fig. 1b). The time for the electrons to attain the steady state is about 0.5 ps.

Figures 1 and 2 show an overshoot in the non-equilibrium temperature of electrons and in the drift velocity of electrons. An analysis of the different channels of pumping and relaxation allows us to conclude that the overshoot at sufficiently high fields is a consequence of the interplay of energy and momentum relaxation times. No overshoot occurs when the momentum relaxation time, which is smaller than the energy relax-ation time shortly after application of the electric field, becomes predominantly larger than the other. On the other hand, the overshoot follows at intermediate to high fields when the relaxation time for energy is constantly larger than the one for momentum. Moreover, it is verified that the peak in velocity follows in the time interval where the drift kinetic energy $m_e^* v_e^2/2$ increases more rapidly than the thermal energy $k_B T_e^*$ and the peak in nonequilibrium temperature follows for a minimum of the quotient of these two energies. For further details on the conditions necessary to occur the phenomenon of overshoot see [53].

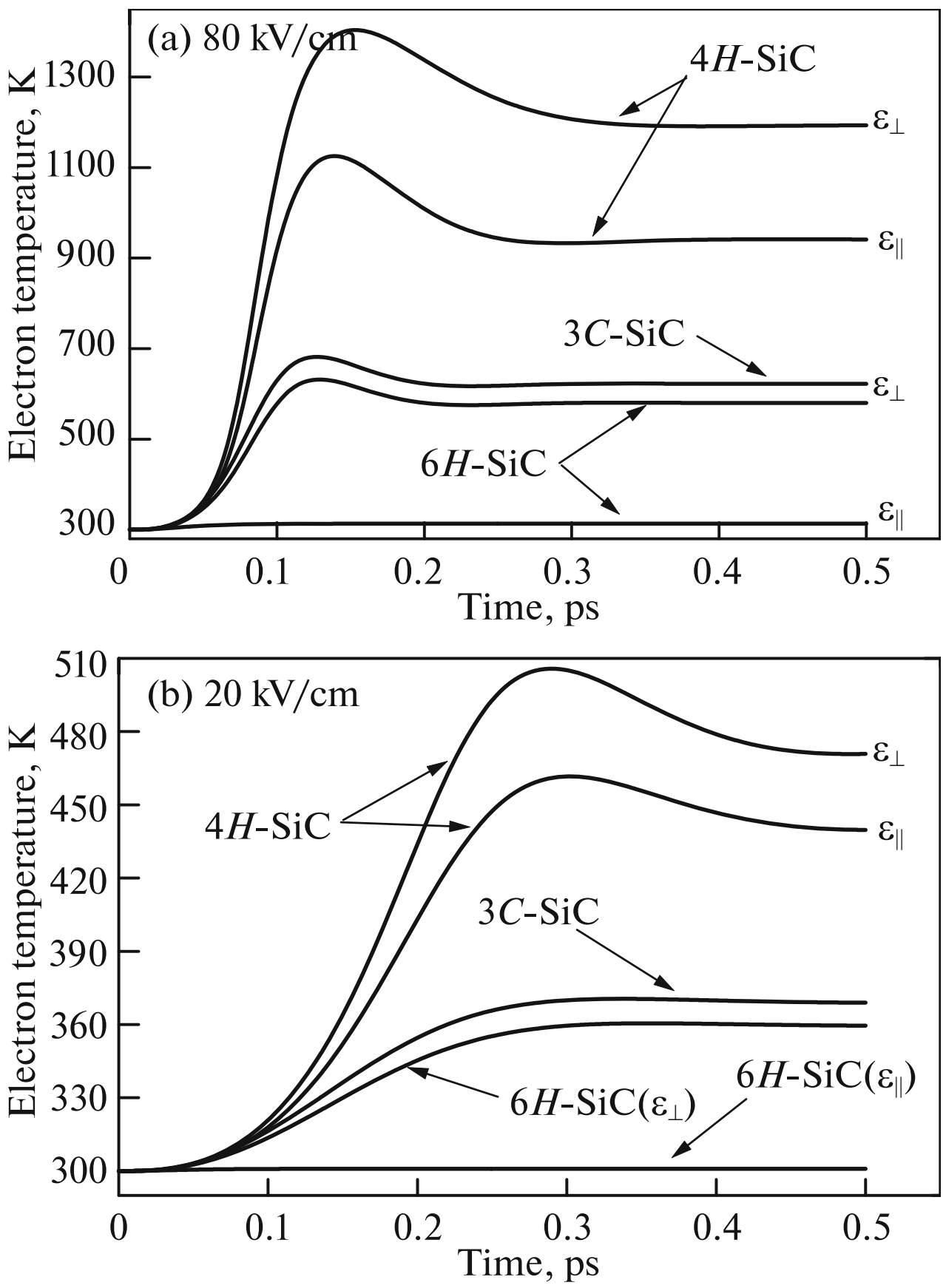

**Fig. 1.** Non-equilibrium temperature of electrons, $T_e^*$, versus time for $n$-type doped SiC-polytypes for an electric field strength of 80 kV/cm (Fig. 1a) and 20 kV/cm (Fig. 1b) with $n = 10^{16}$ cm$^{-3}$.

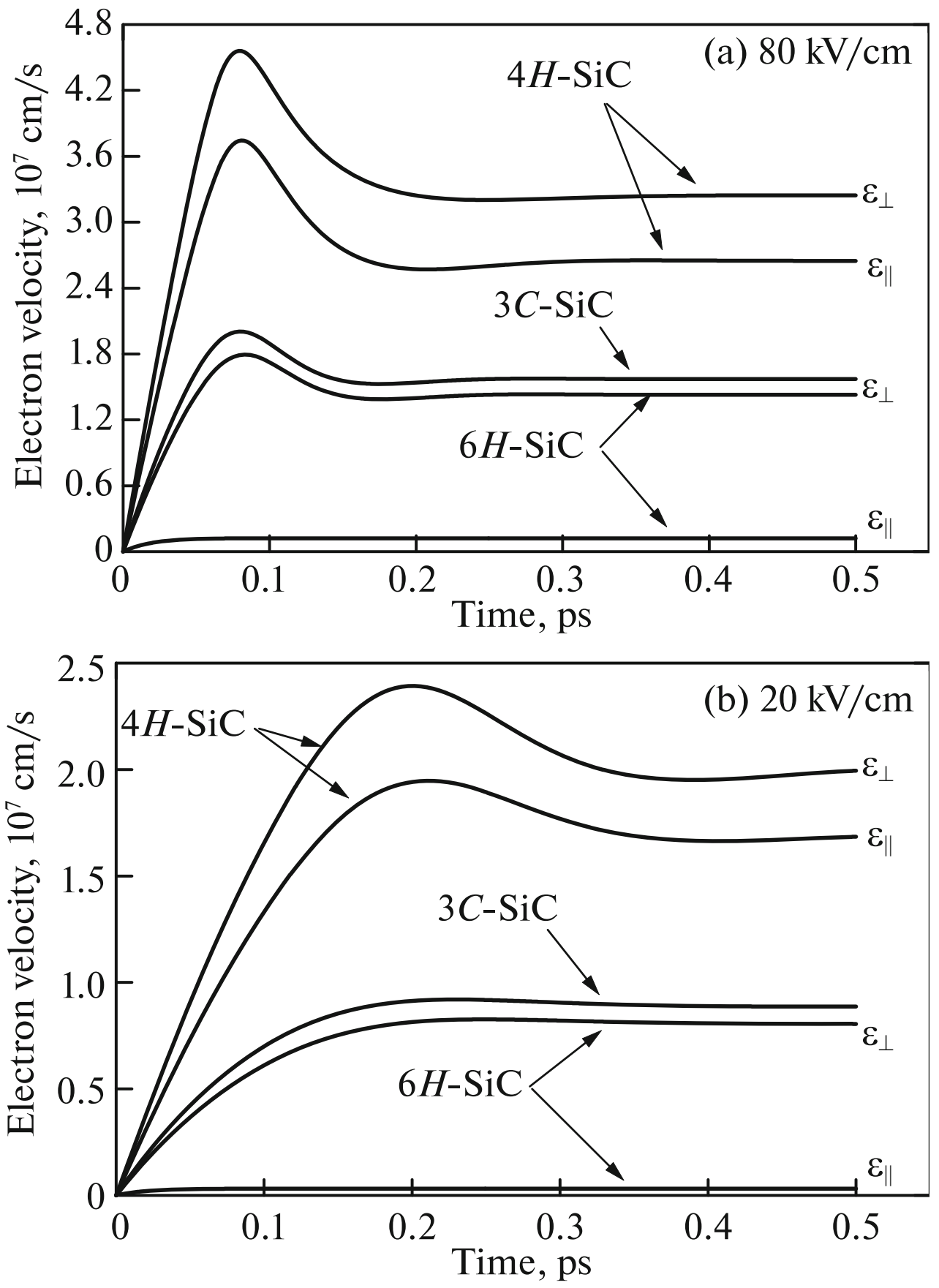

**Fig. 2.** Drift velocity of electrons, $v_e$, versus time for $n$-type doped SiC-polytypes for an electric field strength of 80 kV/cm (Fig. 2a) and 20 kV/cm (Fig. 2b) with $n = 10^{16}$ cm$^{-3}$.

Figure 3 shows the electron distance traveled in $n$-type doped SiC-polytypes for an electric field strength of 80 kV/cm (Fig. 3a) and 20 kV/cm (Fig. 3b). At a time interval of 0.5 ps the values of the distance traveled are in the order of $10^{-2}$ to $10^{-1}$ μm. For the polytype 4$H$-SiC with the electric field applied perpendicular to the $c$-axis ($\mathscr{E}_\perp$), for example, in the electric field intensity of 20 kV/cm the electron distance traveled is 0.09 μm and for 80 kV/cm is 0.16 μm. That is, for an electric field 4 times greater, the distance traveled is only double. This is explained by the nonlinear electron drift velocity behavior as a function of the applied electric field intensity, as shown in Fig. 2 (and also by Fig. 6). For a better visualization of this nonlinear behavior, the Fig. 4 shows the dependence on the electric field strength of the electron distance traveled at a time interval of 0.5 ps.

We did not find in the literature experimental measurements with ultrafast temporal resolution for SiC-polytypes (3$C$-SiC, 4$H$-SiC and 6$H$-SiC) to compare with our theoretical results presented here. However, we can give some suggestions. The carriers' quasitemperature, as well as the drift velocity, can be obtained in experiments measuring optical properties: $T_e^*(t)$ in measurements of luminescence (e.g., see [54]) and $v_e(t)$ in measurements of electroabsorption [55] or with better definition in measurements of Raman scattering by plasmon modes [56].

Figure 5 shows, in the steady state, the increase of non-equilibrium temperature of electrons (in Kelvin) with the increase of applied electric field strength in $n$-type doped SiC-polytypes. This increase in non-equilibrium temperature of electrons comes from the interaction of the carriers with the applied electric field. The increase in non-equilibrium temperature of the longitudinal optical phonons, $T^*_{\rm LO}$, and in the non-equilibrium temperature of acoustical phonons, $T^*_{\rm AC}$, not shown here, was insignificant: less than 1%.

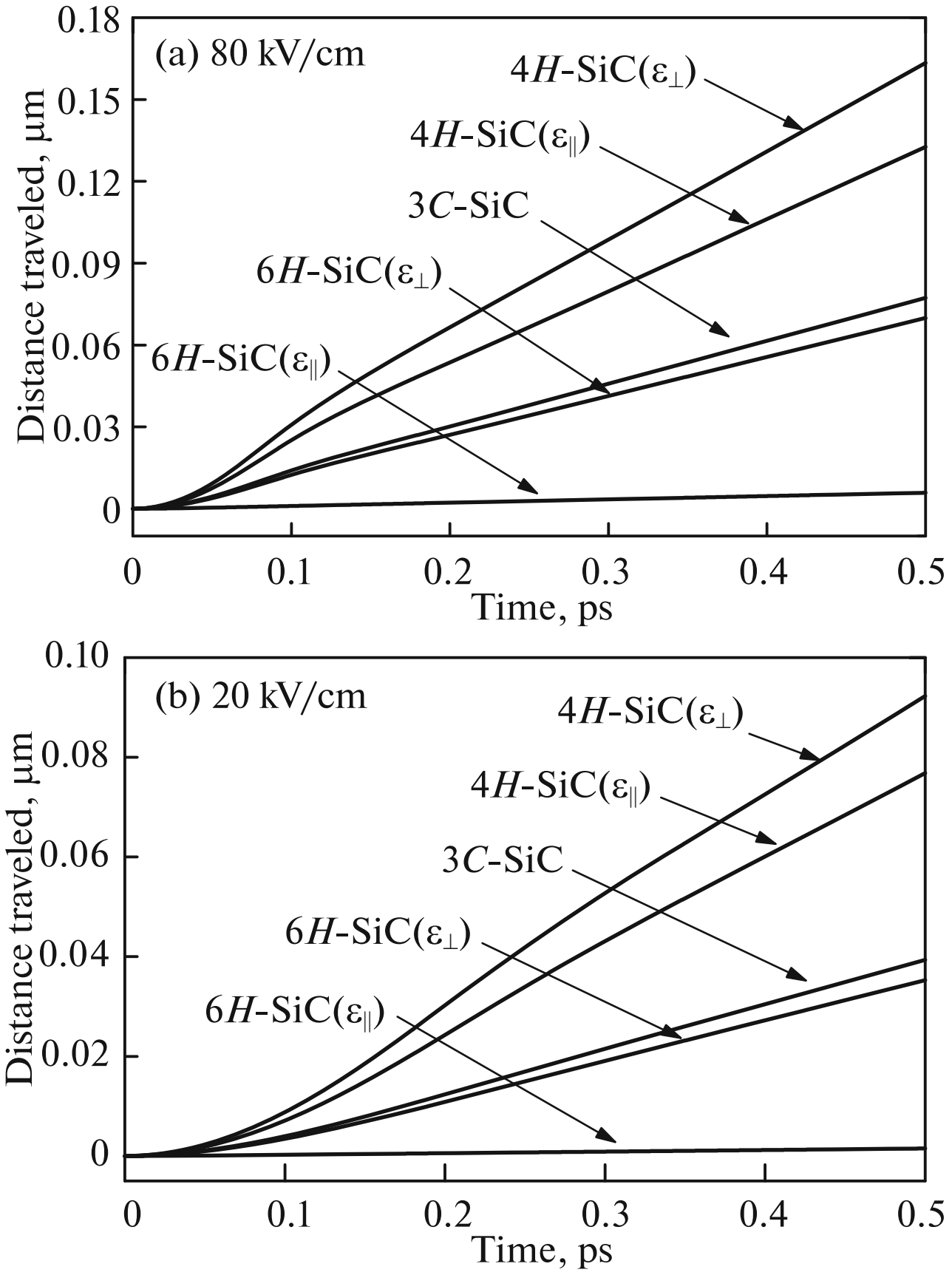

**Fig. 3.** Electron distance traveled in $n$-type doped SiC-polytypes versus time, for an electric field strength of 80 kV/cm (Fig. 3a) and 20 kV/cm (Fig. 3b) with $n = 10^{16}$ cm$^{-3}$.

Figure 6 shows the dependence of the drift velocity of electrons, in the steady state, with the electric field magnitude in $n$-type doped SiC-polytypes. It is noted that there is an ohmic regime for electric fields less than 3 kV/cm. This ohmic behavior is best evidenced by the Fig. 8a, exalting only the range from 0 to 3 kV/cm. Fig. 7 shows the electron mobility ($\mathcal{M} = v_e/\mathcal{E}$) in terms of the electric field strength. In accordance with Fig. 6 at low electric fields, an ohmic region is present, however, limited to approximately 3 kV/cm. Beyond this value of 3 kV/cm there follows a departure from the ohmic behavior. The ohmic region (where mobility is practically constant) is more clearly evidenced by Fig. 8b, exalting only the range from 0 to 3 kV/cm. The change of the electron mobility at higher electric fields is directly related to the non-equilibrium temperature of electrons: with increasing intensity of the electric field the mean energy of the carriers increases, which it is constituted by the kinetic thermal energy (roughly $3k_{\mathrm{B}}T_e^*/2$) and the kinetic energy of drift ($m_e^* v_e^2/2$). The pumped energy (from the applied electric field) is distributed as to privilege the thermal energy (thermal-chaotic movement) instead of the (organized) kinetic energy of drift.

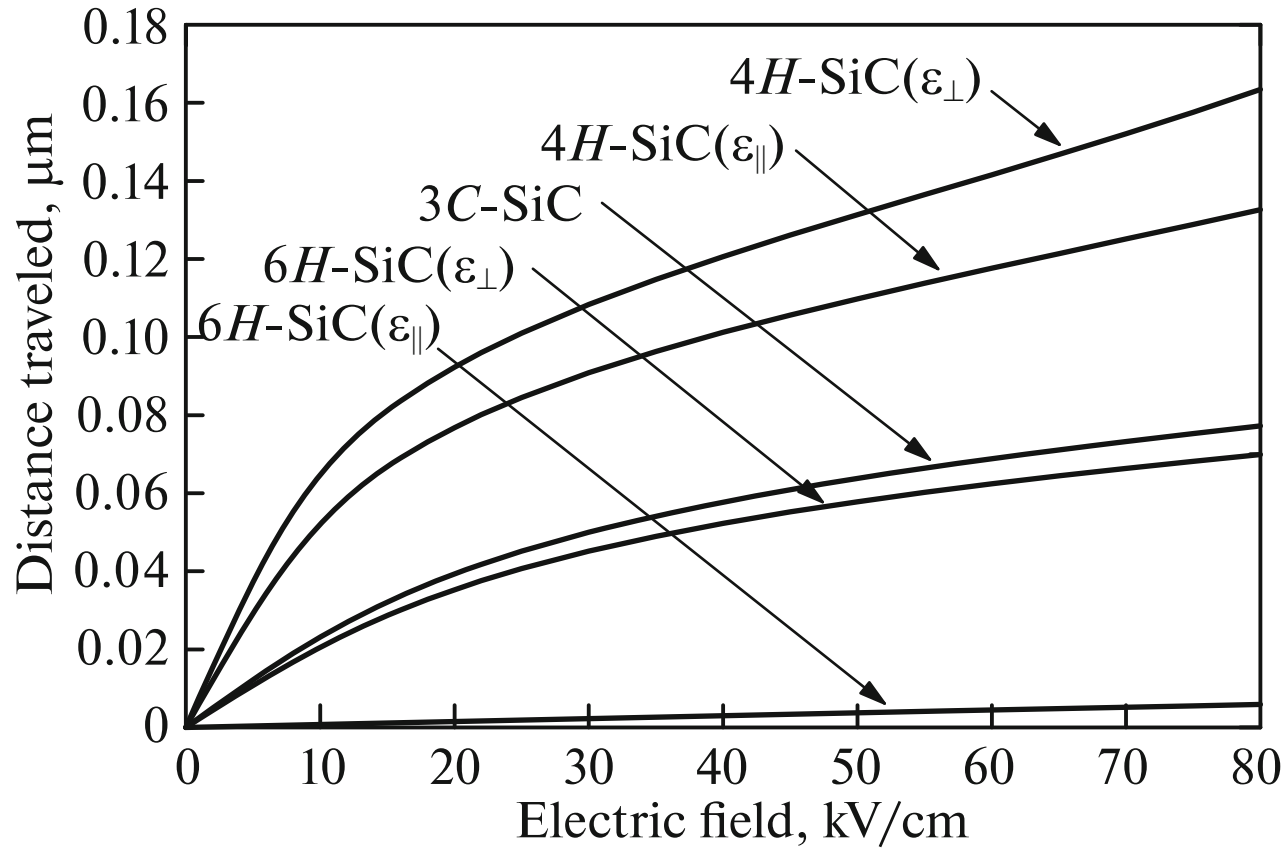

**Fig. 4.** Electron distance traveled in $n$-type doped SiC-polytypes, at a time interval of 0.5 ps, versus electric field strength, with $n = 10^{16}$ cm$^{-3}$.

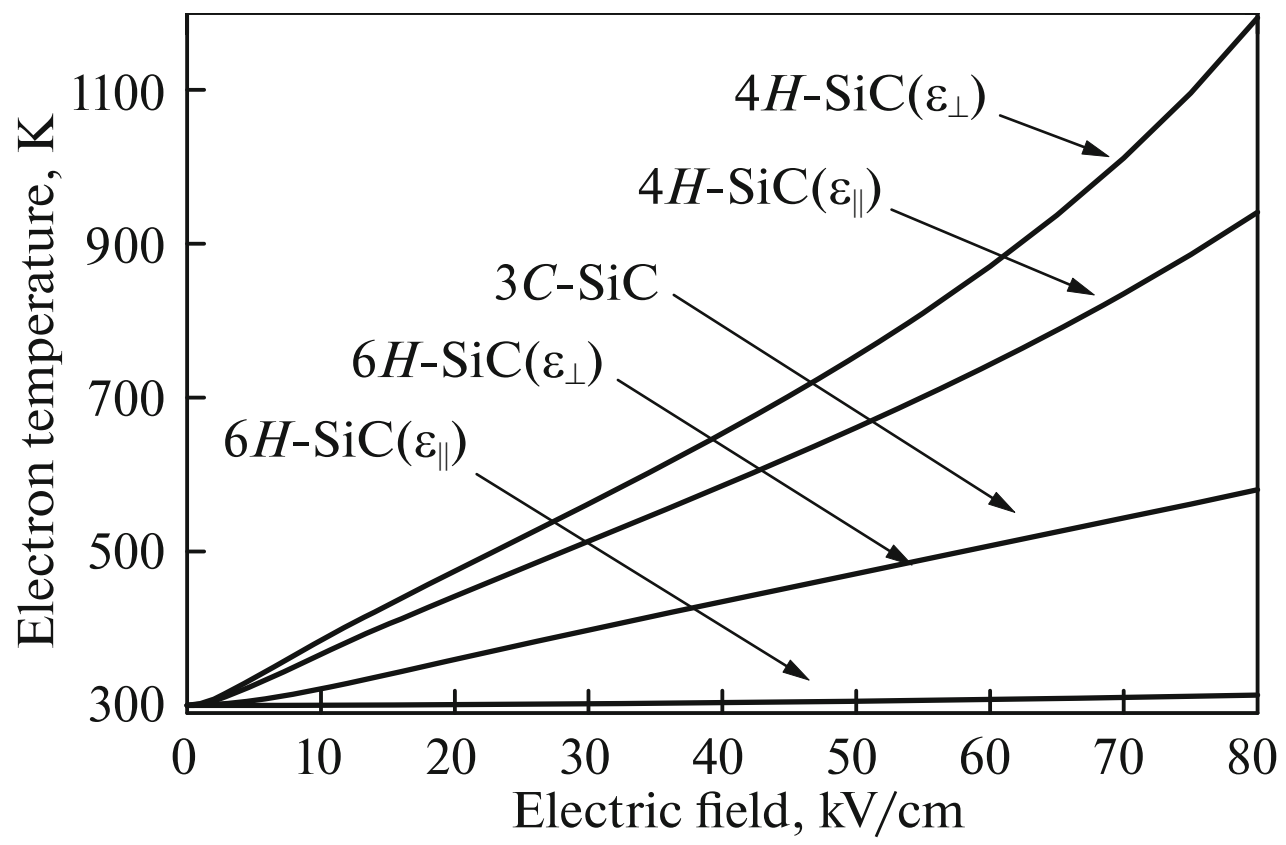

**Fig. 5.** Non-equilibrium temperature of electrons, $T_e^*$, versus electric field magnitude in the steady state of $n$-type doped SiC-polytypes, with $n = 10^{16}$ cm$^{-3}$.

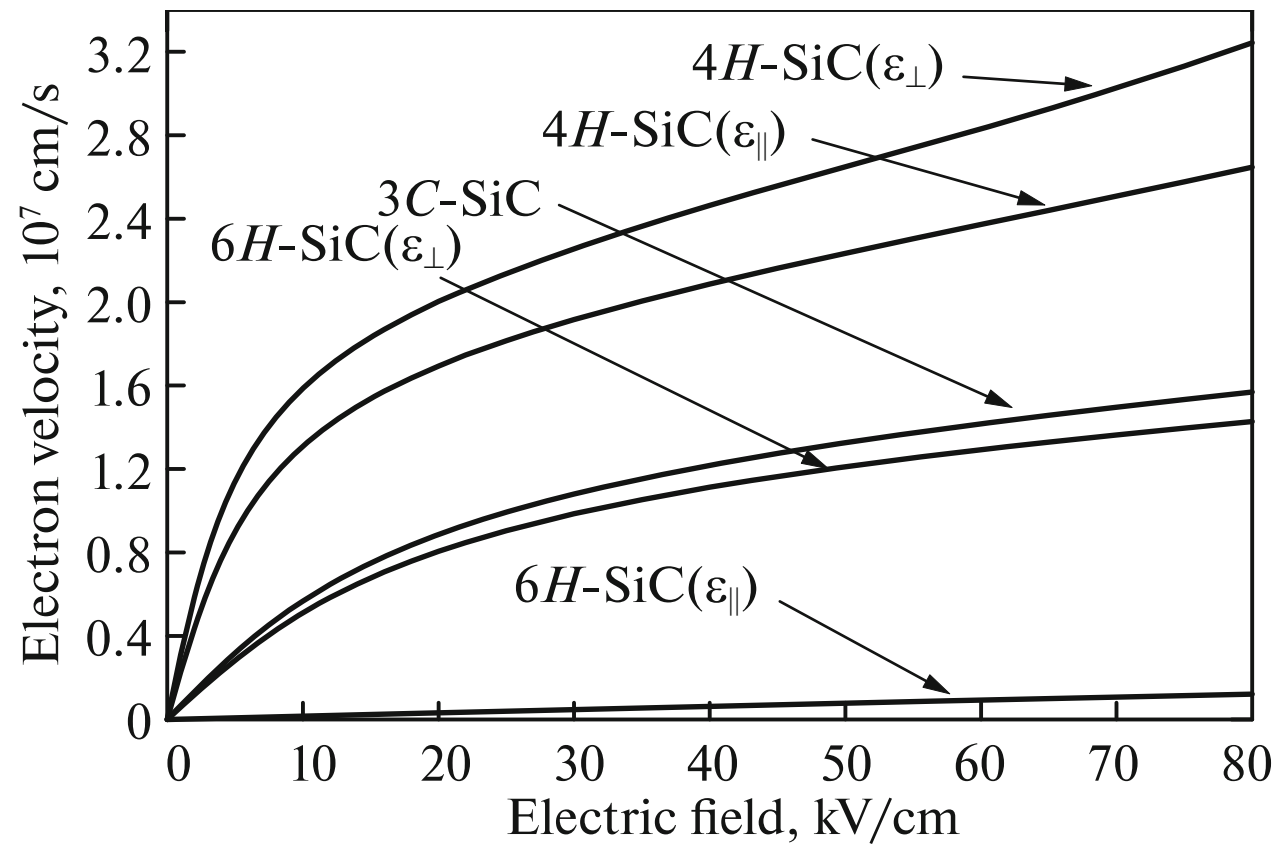

**Fig. 6.** Drift velocity of electrons, $v_e$, versus electric field magnitude in the steady state of $n$-type doped SiC-polytypes, with $n = 10^{16}$ cm$^{-3}$.

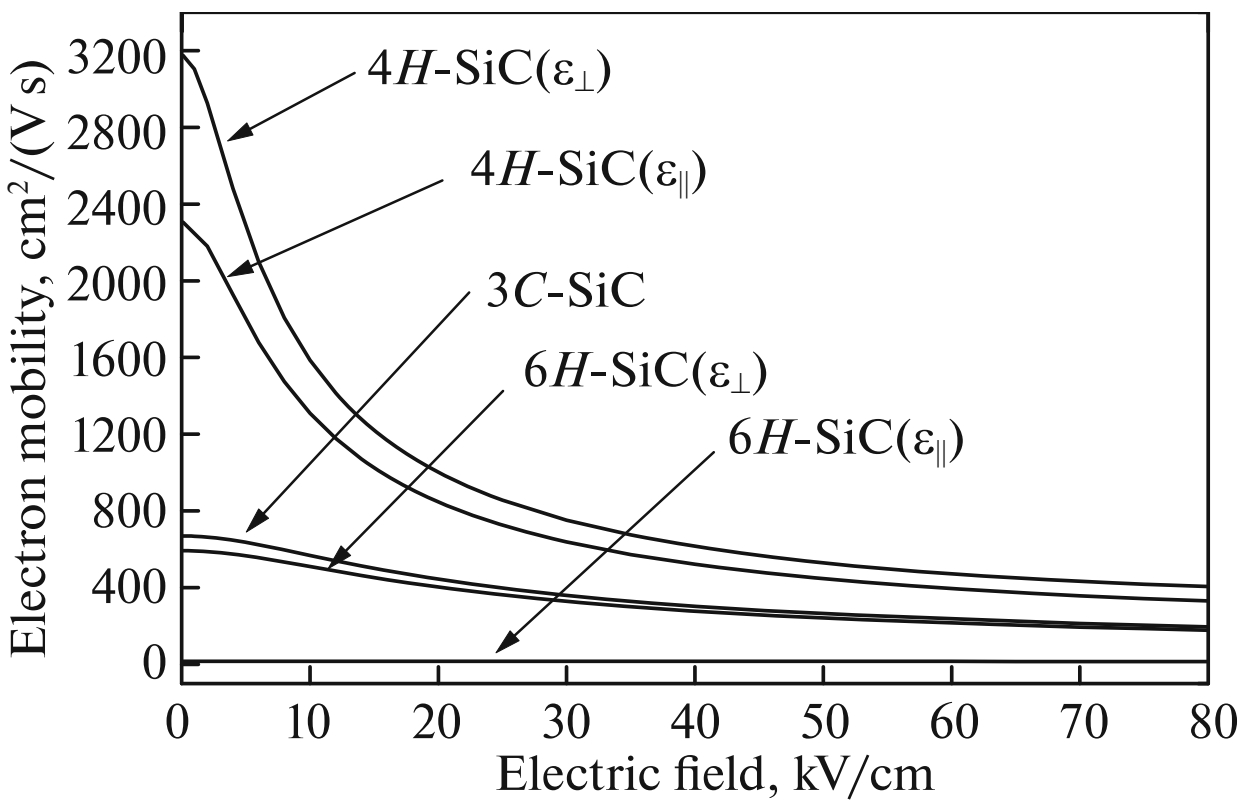

**Fig. 7.** Electron mobility versus electric field magnitude in the steady state of $n$-type doped SiC-polytypes, with $n = 10^{16}$ cm$^{-3}$.

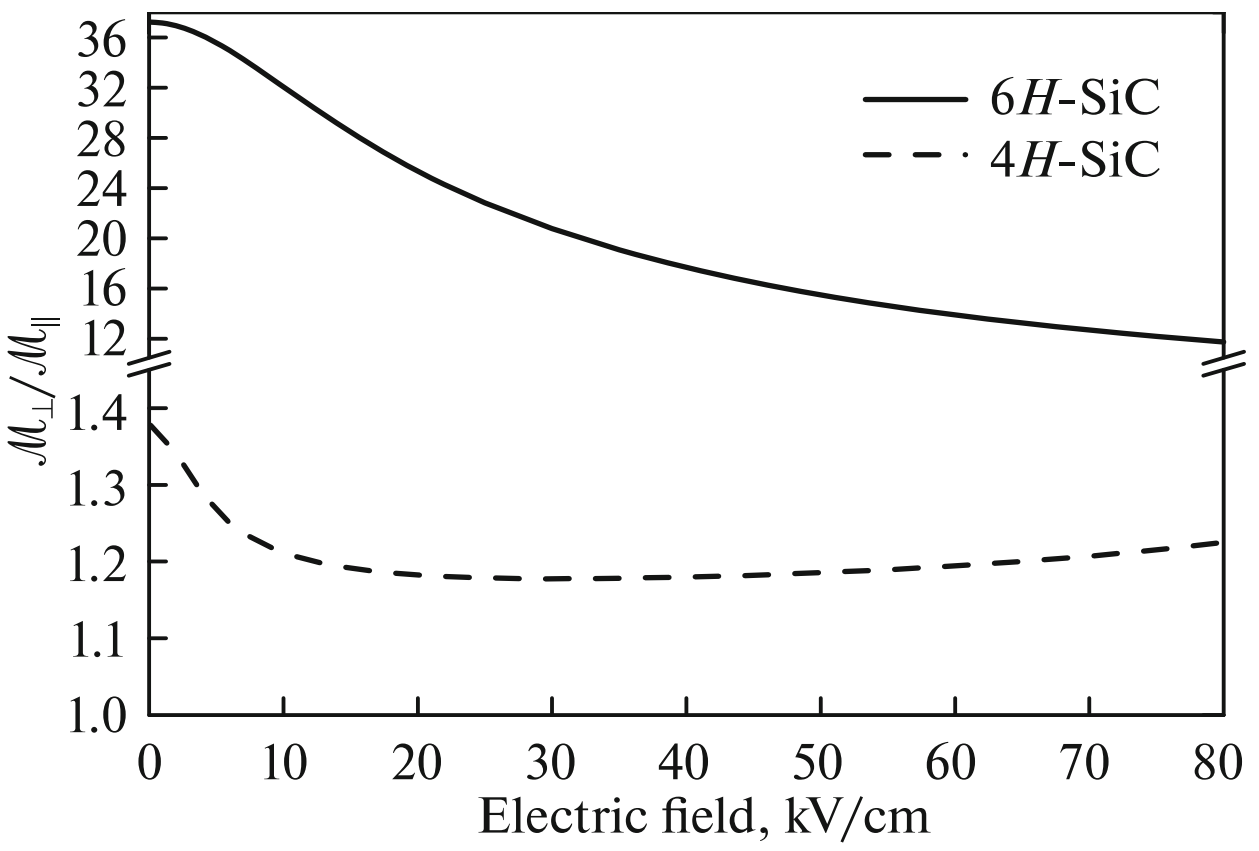

**Fig. 9.** Dependence on the electric field (in steady state) of the ratio $\mathcal{M}_{\perp}/\mathcal{M}_{\parallel}$, with $n = 10^{16}$ cm$^{-3}$.

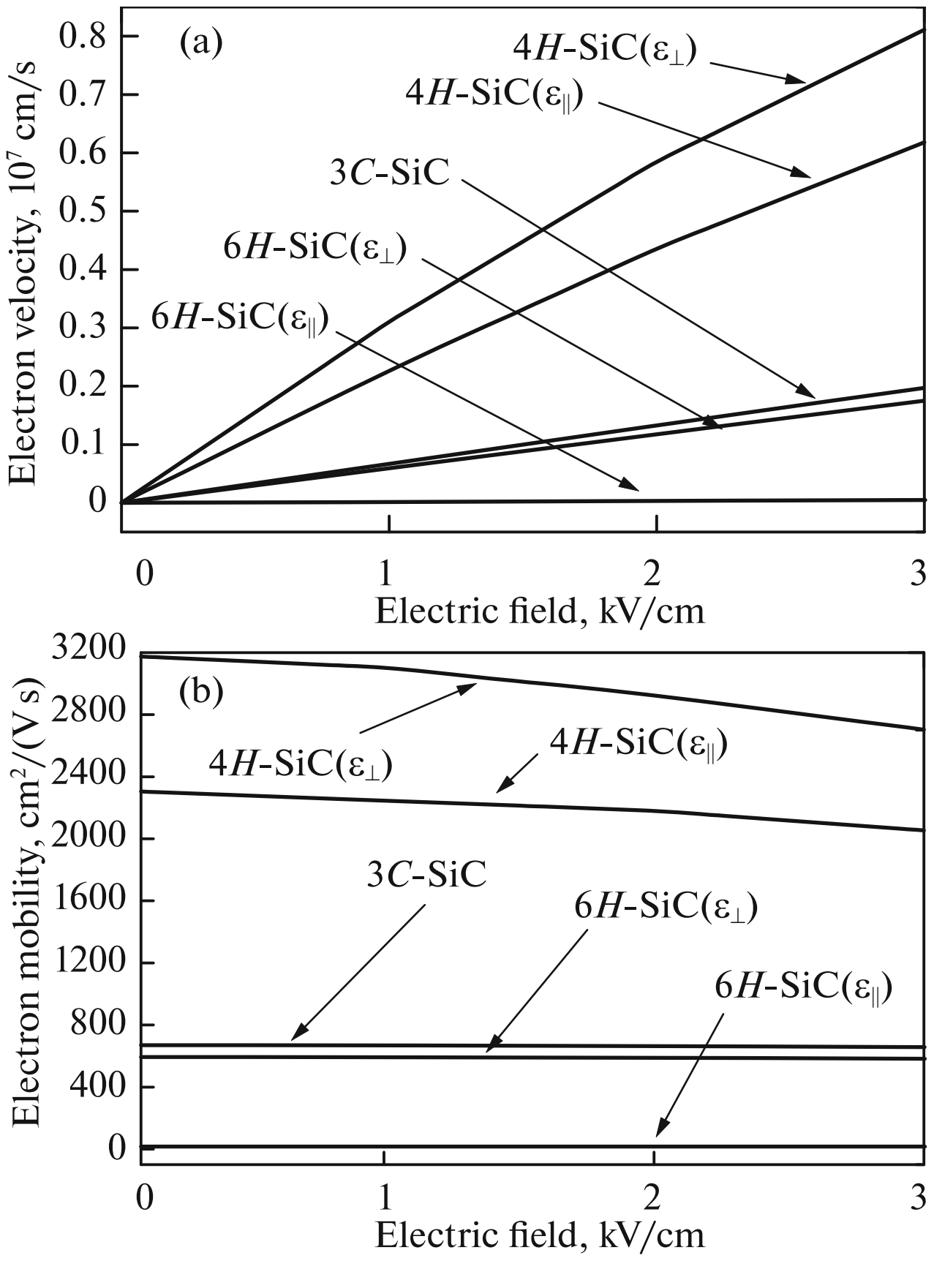

**Fig. 8.** Dependence on the electric field, in steady state, of the drift velocity, figure (a), and electron mobility, figure (b) for low electric field strength of $n$-type doped SiC-polytypes, with $n = 10^{16}$ cm$^{-3}$.

We emphasize that the experimental electron mobilities vary greatly from sample to sample, intimately related to the quality and growth conditions [57]. According to F. Meng et al. [57], despite the abundance data in literature (both experimental and theoretical) the electrical transport properties of SiC-polytypes have not yet been fully understood at the microscopic level.

Figure 9 shows the ratio between the electron mobility with the electric field applied perpendicular to the $c$-axis, $\mathcal{M}(\mathcal{E}_{\perp}) \equiv \mathcal{M}_{\perp}$, and the electron mobility with the electric field applied parallel to the $c$-axis, $\mathcal{M}(\mathcal{E}_{\parallel}) \equiv \mathcal{M}_{\parallel}$, that is: $\mathcal{M}_{\perp}/\mathcal{M}_{\parallel}$. Figure 9 evidences that 6$H$-SiC is very more anisotropic than 4$H$-SiC. Inspection of Fig. 9 tells us that this ratio $\mathcal{M}_{\perp}/\mathcal{M}_{\parallel}$ remains in a range between 11.7 to 37.2 for 6$H$-SiC (solid line) while for 4$H$-SiC (dashed line) in a range between 1.18 to 1.38. Noticed that the ratio $m_{e\parallel}^*/m_{e\perp}^*$ listed in Table 1 is 5.5 for 6$H$-SiC and 1.22 for 4$H$-SiC.

## 4. FINAL COMMENTS

Summarizing, this communication presents a detailed study on the nonlinear charge transport (in both, transient and steady state) in bulk $n$-type doped SiC-polytypes (3$C$-SiC, 4$H$-SiC and 6$H$-SiC) using a Non-Equilibrium Quantum Kinetic Theory derived from the method of Nonequilibrium Statistical Operator (NSO). The region with ohmic and non-ohmic behavior in the electron drift velocity (and electron mobility) dependence on the electric field strength was characterized. According to Table 1, the parameters used in the calculations for the polytypes 3$C$-SiC, 4$H$-SiC and 6$H$-SiC are almost identical, except the electron effective mass. This implies that the different transport properties of 3*C*-SiC, 4*H*-SiC and 6*H*-SiC are solely dominated by their electron effective mass tensors. From the results obtained in this paper, the most attractive of these semiconductors for applications requiring greater electronic mobility is the polytype 4*H*-SiC with the electric field applied perpendicular to the *c*-axis. Summarizing for the *n*-type doped SiC-polytypes studied here, the electron mobility, $\mathcal{M}$, follow the order: $\mathcal{M}_{(4H)}(\mathcal{E}_{\perp}) > \mathcal{M}_{(4H)}(\mathcal{E}_{\parallel}) > \mathcal{M}_{(3C)} > \mathcal{M}_{(6H)}(\mathcal{E}_{\perp}) > \mathcal{M}_{(6H)}(\mathcal{E}_{\parallel})$.

## *APPENDIX A*

## COLLISION TERMS IN EQS. (6)–(9)

The detailed expressions for the collision operators in Eqs. (6)–(9) are:

$$J_{\mathbf{P}_{\rm ph}}(t) = \sum_{\mathbf{k},\mathbf{q},\ell,\eta} \frac{2\pi}{\hbar} \hbar\mathbf{q} |M_{\eta}^{\ell}(\mathbf{q})|^2 [\nu_{\mathbf{q},\eta}(t) f_{\mathbf{k}}(t) \times (1 - f_{\mathbf{k}+\mathbf{q}}(t)) - f_{\mathbf{k}+\mathbf{q}}(t)(1 + \nu_{\mathbf{q},\eta}(t)) \times (1 - f_{\mathbf{k}}(t))] \delta(\epsilon_{\mathbf{k}+\mathbf{q}} - \epsilon_{\mathbf{k}} - \hbar\omega_{\mathbf{q},\eta}),$$

$$\mathbf{J}_{\mathbf{P}_{\rm imp}}(t) = -\frac{128\sqrt{2\pi m_e^*}(k_B T_e^*(t))^{3/2}}{n_I (\mathcal{Z}e^2/\epsilon_0)^2 G(t)} \mathbf{P}_e(t),$$

$$J_{E,\rm ph}(t) = \sum_{\mathbf{k},\mathbf{q},\ell,\eta} \frac{2\pi}{\hbar} |M_{\eta}^{\ell}(\mathbf{q})|^2 (\epsilon_{\mathbf{k}+\mathbf{q}} - \epsilon_{\mathbf{k}}) [\nu_{\mathbf{q},\eta}(t) \times f_{\mathbf{k}}(t)(1 - f_{\mathbf{k}+\mathbf{q}}(t)) - f_{\mathbf{k}+\mathbf{q}}(t) \times (1 + \nu_{\mathbf{q},\eta}(t))(1 - f_{\mathbf{k}}(t))] \times \delta(\epsilon_{\mathbf{k}+\mathbf{q}} - \epsilon_{\mathbf{k}} - \hbar\omega_{\mathbf{q},\eta}),$$

$$J_{{\rm LO},an}(t) = -\sum_{\mathbf{q}} \hbar\omega_{\mathbf{q},{\rm LO}} \frac{\nu_{\mathbf{q},{\rm LO}}(t) - \nu_{\mathbf{q},{\rm AC}}(t)}{\tau_{{\rm LO},an}},$$

$$J_{{\rm AC},dif}(t) = -\sum_{\mathbf{q}} \hbar\omega_{\mathbf{q},{\rm AC}} \frac{\nu_{\mathbf{q},{\rm AC}}(t) - \nu_0}{\tau_{ac,dif}}.$$

The quantities $M_{\eta}^{\ell}(\mathbf{q})$ are the matrix elements of the interaction between carriers and branch $\eta$-type phonons ($\eta$ = LO, AC for longitudinal optical and acoustical phonons, respectively), with supraindex $\ell$ indicating the kind of interaction (polar, deformation potential, piezoelectric, etc.); $\nu_{\mathbf{q},\eta}(t)$ and $f_{\mathbf{k}}(t)$ are, respectively, the phonons and electrons distribution; $\delta$ is delta function and $\epsilon_{\mathbf{k}} = \hbar^2 k^2/2m_e^*$; $n_I$ is the density of impurities, $\mathcal{L}$ the units of charge of the impurity, and

$$G(t) = \ln(1 + b(t)) - \frac{b(t)}{1 + b(t)},$$

where $b(t) = 24\epsilon_0 m_e^* [k_B T_e^*(t)]^2 / n_I e^2 \hbar^2$. Finishing, $\tau_{{\rm LO},an}$ is a relaxation time which is obtained from the band width in Raman scattering experiments [58], and $\tau_{{\rm AC},dif}$ is a characteristic time for heat diffusion, which depends on the particularities of the contact of sample and reservoir [59]. We notice that more details for the collision operators are given in [32].